\documentclass[prx,aps,twocolumn,floatfix,showpacs,superscriptaddress]{revtex4-1}
\usepackage{amssymb}
\usepackage{amsmath}
\usepackage{graphicx}
\usepackage[colorlinks=true,linkcolor=blue,anchorcolor=red,citecolor=blue,urlcolor=blue]{hyperref}

\begin{document}

\title{High-field magnetoconductivity of topological semimetals with short-range
potential}

\author{Hai-Zhou Lu}

\affiliation{Department of Physics, South University of Science and Technology
of China, Shenzhen, China}

\affiliation{Department of Physics, The University of Hong Kong, Pokfulam Road,
Hong Kong, China}

\author{Song-Bo Zhang}

\affiliation{Department of Physics, The University of Hong Kong, Pokfulam Road,
Hong Kong, China}

\author{Shun-Qing Shen}

\affiliation{Department of Physics, The University of Hong Kong, Pokfulam Road,
Hong Kong, China}

\date{\today }
\begin{abstract}
Weyl semimetals are three-dimensional topological states of matter,
in a sense that they host paired monopoles and antimonopoles of Berry
curvature in momentum space, leading to the chiral anomaly. The chiral
anomaly has long been believed to give a positive magnetoconductivity
or negative magnetoresistivity in strong and parallel fields. However,
several recent experiments on both Weyl and Dirac topological semimetals
show a negative magnetoconductivity in high fields. Here, we study
the magnetoconductivity of Weyl and Dirac semimetals in the presence
of short-range scattering potentials. In a strong magnetic field applied
along the direction that connects two Weyl nodes, we find that the
conductivity along the field direction is determined by the Fermi
velocity, instead of by the Landau degeneracy. We identify three scenarios
in which the high-field magnetoconductivity is negative. Our findings
show that the high-field positive magnetoconductivity may not be a
compelling signature of the chiral anomaly and will be helpful for
interpreting the inconsistency in the recent experiments and earlier
theories.
\end{abstract}


\maketitle

\section{Introduction}

\begin{figure}[tbph]
\centering \includegraphics[width=0.12\textwidth]{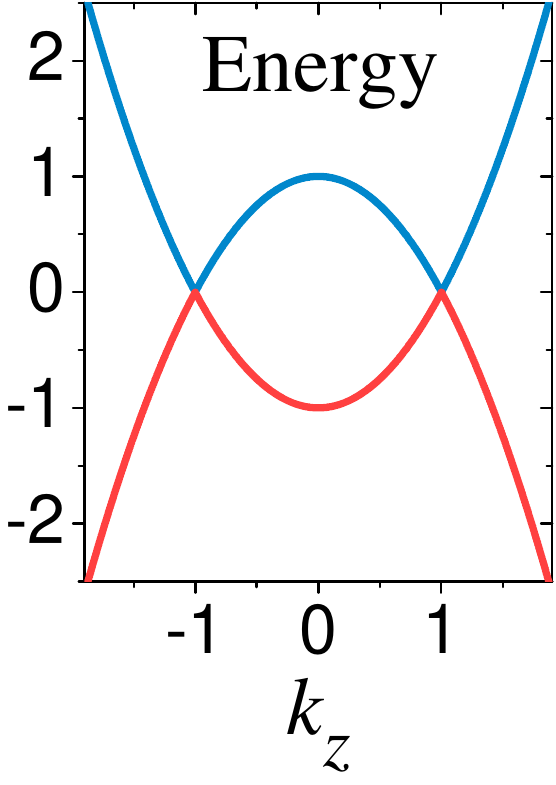}
\includegraphics[width=0.35\textwidth]{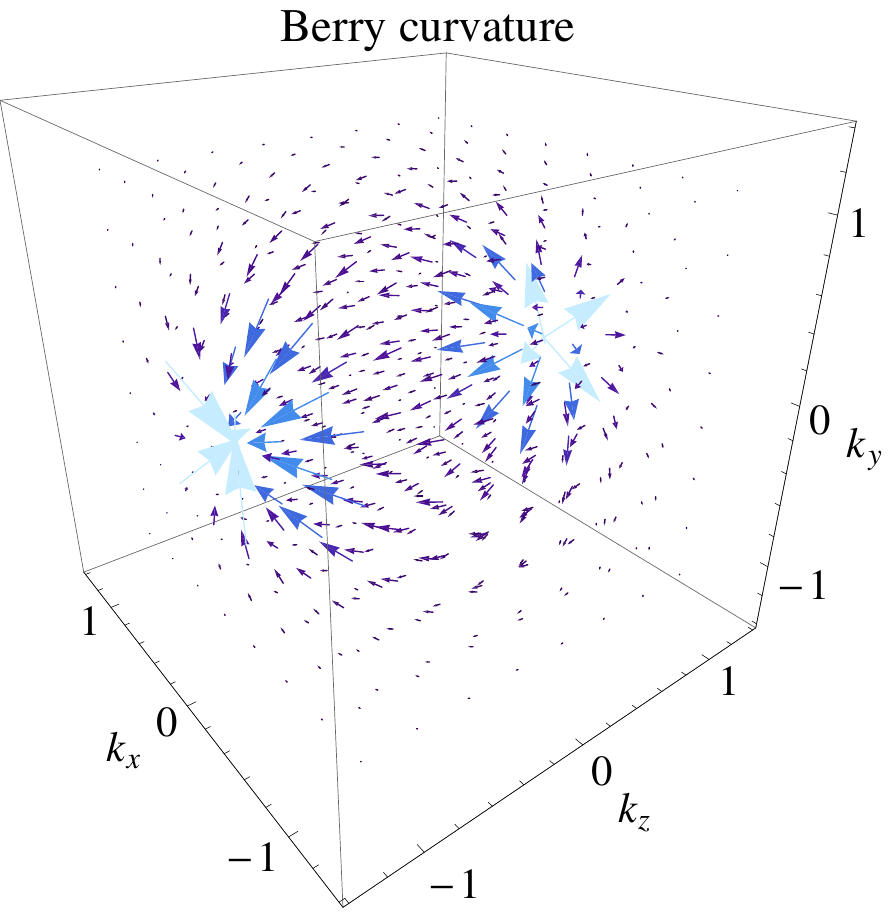}
\caption{Left: the energy spectrum of a Weyl semimetal as a function of $k_{z}$
at $k_{x}=k_{y}=0$. Right: the vector plot of the Berry curvature
of a Weyl semimetal. $(k_{x},k_{y},k_{z})$ is the wave vector. The
analytic expression of the three-dimensional Berry curvature is given
in Eq. (\ref{Eq:Berry}). The parameters are $M_{0}=M_{1}=A=1$, so
$k_{c}=1$. }
\label{Fig:Berry}
\end{figure}

Topological semimetals are three-dimensional topological states of
matter. Their band structures look like three-dimensional analogue
of graphene, in which the conduction and valence energy bands with
linear dispersions touch at a finite number of points, i.e., Weyl nodes
\cite{Balents11physics}. The nodes always occur in pairs and carry
opposite chirality. One of the topological aspects of Weyl semimetals
is that they host pairs of monopole and anti-monopole of Berry curvature
in momentum space \cite{Volovik-book,Wan11prb} (see Fig.~\ref{Fig:Berry}),
and the fluxes of Berry curvature flow from one monopole to the other.
In the presence of both a magnetic field and an electric field along
the direction that connects two monopoles, electrons can be pumped
from one monopole to the other, leading to the Adler-Bell-Jackiw chiral
anomaly \cite{Adler69pr,Bell69Jackiw,Nielsen81npb} (also known as
triangle anomaly). Recently, angle-resolved photoemission spectroscopy
(ARPES) has identified the Dirac nodes \cite{Young12prl} (doubly-degenerate
Weyl nodes) in (Bi$_{1-x}$In$_{x}$)$_{2}$Se$_{3}$ \cite{Brahlek12prl,Wu13natphys},
Na$_{3}$Bi \cite{Wang12prb,Liu14sci,Wang13prb,Xu15sci}, and Cd$_{3}$As$_{2}$
\cite{Wang13prb,Liu14natmat,Neupane14nc,Yi14srep,Borisenko14prl}
and Weyl nodes in TaAs \cite{Weng15prx,Huang15arXiv,Lv15arXiv,Xu15arXiv}.
Also, scanning tunneling microscopy has observed the Landau quantization
in Cd$_{3}$As$_{2}$ \cite{Jeon14natmat} and TlBiSSe \cite{Novak14arXiv}.

While the chiral anomaly is well established in momentum space, it
becomes a challenging issue how to detect the effects of the chiral
anomaly or relevant physical consequences. This has been attracting
a lot of theoretical efforts, such as the prediction of negative parallel
magnetoresistance \cite{Nielsen83plb,Aji12prbr,Son13prb,Burkov14prl-chiral},
proposal of non-local transport \cite{Parameswaran14prx}, using electronic
circuits \cite{Kharzeev13prb}, plasmon mode \cite{ZhouJH15prb},
etc. In particular, whether or not the chiral anomaly could produce
measurable magnetoconductivity is one of the focuses in recent efforts.
This has inspired a number of transport experiments in topological
semimetals Cd$_{3}$As$_{2}$ \cite{Liang15nmat,Feng14arXiv,He14prl,Zhao14arXiv,Cao14arXiv,Narayanan15prl}, ZrTe$_5$ \cite{Li14arXiv}, 
NbP \cite{Shekhar15arXiv}, Na$_{3}$Bi \cite{Xiong15arXiv}, and
TaAs \cite{HuangXC15arXiv,ZhangCL15arXiv}. The chiral anomaly has
been claimed to be verified in several different topological semimetals,
including BiSb alloy \cite{Kim13prl}, ZrTe$_5$ \cite{Li14arXiv}, TaAs \cite{HuangXC15arXiv,ZhangCL15arXiv},
and Na$_{3}$Bi \cite{Xiong15arXiv}, in which similar magnetoconductivity behaviors are
observed when the magnetic field is applied along the conductivity
measurement direction: (i) In weak fields, a $-\sqrt{B}$ negative
magnetoconductivity is observed at low temperatures, consistent with
the quantum transport theory of the weak antilocalization of Weyl
or Dirac fermions in three dimensions \cite{Kim13prl,Lu14arXiv}. (ii)
In intermediate fields, a $B^{2}$ positive magnetoconductivity is
observed, as expected by the theory of the semiclassical conductivity
arising from the chiral anomaly \cite{Son13prb,Ramakrishnan15arXiv,Burkov14prl-chiral}.
(iii) In high fields, the magnetoconductivity is always negative in
the experiments. However, in the strong-field limit, a positive magnetoconductivity
is expected in existing theories, also as one of the signatures of
the chiral anomaly \cite{Nielsen83plb,Aji12prbr,Son13prb,Gorbar14prb}.

In this work, we focus on the high-field limit, and present a systematic
calculation on the conductivity of topological semimetals. Beyond
the previous treatments, we start with a two-node model that describes
a pair of Weyl nodes with a finite distance in momentum space. Moreover,
we fully consider the magnetic field dependence of the scattering
time for electrons on each Landau level, and obtain a conductivity
formula. The efforts lead to qualitatively distinct results compared
to all the previous theories. We find that the conductivity does not
grow with the Landau degeneracy but mainly depends on the Fermi velocity.
The magnetoconductivity arises from the field dependence of the Fermi
velocity when the chemical potential is tuned by the magnetic field.
Based on this formula and the model, we find that although the positive
magnetoconductivity is also possible, three cases can be identified in which
the magnetoconductivity is negative, possibly applicable to those
observed in the experiments in high magnetic fields.

The paper is organized as follows. In Sec. \ref{sec:model}, we introduce
the two-node model and show how it carries all the topological properties
of a topological semimetal. In Sec. \ref{sec:Landau}, we present
the solutions of the Landau bands of the semimetal in a magnetic field
applied along the $z$ direction (the two Weyl nodes are separated
along this direction). In Sec. \ref{sec:sigma-z}, we calculate the
$z$-direction magnetoconductivity in the presence of the short-range
delta scattering potential. In Sec. \ref{sec:scenarios}, we discuss
various scenarios that the negative or positive magnetoconductivity
may occur. In Sec. \ref{sec:sigma-x}, we present the transport in
the $x-y$ plane, including the $x$-direction conductivity and the
Hall conductance. Finally, a summary of three scenarios of the negative
magnetoconductivity is given in Sec. \ref{sec:summary}. The details
of the calculations are provided in Appendices \ref{sec:Lambda}-\ref{sigma-01-app}.

\section{Model and its topology}

\label{sec:model}

A minimal model for a Weyl semimetal is
\begin{eqnarray}\label{eq:model}
H=A(k_{x}\sigma_{x}+k_{y}\sigma_{y})+\mathcal{M_{\mathbf{k}}}\sigma_{z},
\end{eqnarray}
where $\sigma$ are the Pauli matrices, $\mathcal{M}_{\mathbf{k}}=M_{0}-M_{1}(k_{x}^{2}+k_{y}^{2}+k_{z}^{2})$,
$(k_{x},k_{y},k_{z})$ is the wave vector, and $A$, $M_{0/1}$ are
model parameters. This minimal model gives a global description of
a pair of Weyl nodes of opposite chirality and all the topological
properties. It has an identical structure as that for A-phase of $^{3}$He
superfluids \cite{Shen-book} The dispersions of two energy bands of
this model are $E_{\pm}=\pm\sqrt{\mathcal{M}_{\mathbf{k}}^{2}+A^{2}(k_{x}^{2}+k_{y}^{2})}$,
which reduce to $E_{\pm}=\pm|M_{0}-M_{1}k_{z}^{2}|$ at $k_{x}=k_{y}=0$.
If $M_{0}M_{1}>0$, the two bands intersect at $(0,0,\pm k_{c})$
with $k_{c}\equiv\sqrt{M_{0}/M_{1}}$ (see Fig.~\ref{Fig:Berry}).
Around the two nodes $(0,0,\pm k_{c})$, $H$ reduces to two separate
local models
\begin{equation}
H_{\pm}=\mathbf{M}_{\pm}\cdot\sigma,
\end{equation}
$H_{\pm}=\mathbf{M}_{\pm}\cdot\sigma$with $\mathbf{M}_{\pm}=\left(A\widetilde{k}_{x},A\widetilde{k}_{y},\mp2M_{1}k_{c}\widetilde{k}_{z}\right)$
and $(\widetilde{k}_{x},\widetilde{k}_{y},\widetilde{k}_{z})$ the
effective wave vector measured from the Weyl nodes.

The topological properties in $H$ can be seen from the Berry curvature
\cite{Xiao10rmp}, $\Omega(\mathbf{k})$ = $\nabla_{\mathbf{k}}\times\mathbf{A}(\mathbf{k})$,
where the Berry connection is defined as $\mathbf{A}(\mathbf{k})$
= $i\left\langle u(\mathbf{k})\right|\nabla_{\mathbf{k}}\left|u(\mathbf{k})\right\rangle $.
For example, for the energy eigenstates for the $+$ band $\left|u(\mathbf{k})\right\rangle $
= $[\cos(\Theta/2),\sin(\Theta/2)e^{i\varphi}]$, where $\cos\Theta\equiv\mathcal{M}_{\mathbf{k}}/E_{+}$
and $\tan\varphi\equiv k_{y}/k_{x}$. The three-dimensional Berry
curvature for the two-node model can be expressed as
\begin{eqnarray}
\boldsymbol{\Omega}\left(\mathbf{k}\right)=\frac{A^{2}M_{1}}{E_{+}^{3}}\left[k_{z}k_{x},k_{z}k_{y},\frac{1}{2}\left(k_{z}^{2}-k_{c}^{2}-k_{x}^{2}-k_{y}^{2}\right)\right].\label{Eq:Berry}
\end{eqnarray}
When $M_{0}M_{1}>0$, there exist a pair of singularities at $(0,0,\pm k_{c})$
as shown in Fig. {\ref{Fig:Berry}}. The chirality of a Weyl node
can be found as an integral over the Fermi surface enclosing one Weyl
node $(1/2\pi)\oint\Omega(\mathbf{k})\cdot d\mathbf{S(\mathbf{k})}$,
which yields opposite topological charges $\mp\mathrm{sgn}(M_{1})$
at $\pm k_{c}$, corresponding to a pair of ``magnetic monopole and
antimonopole'' in momentum space. For a given $k_{z}$, a Chern number
can be well defined as $n_{c}(k_{z})=-(1/2\pi)\iint dk_{x}dk_{y}\Omega(\mathbf{k})\cdot\hat{z}$
to characterize the topological property in the $k_{x}$-$k_{y}$
plane, and $n_{c}(k_{z})=-\frac{1}{2}[\mathrm{sgn}(M_{0}-M_{1}k_{z}^{2})+\mathrm{sgn}(M_{1})]$
\cite{Lu10prb}. For $M_{0}M_{1}>0$, $n_{c}(k_{z})=-\mathrm{sgn}(M_{1})$
for $-k_{c}<k_{z}<k_{c}$, and $n_{c}(k_{z})=0$ for other cases \cite{Yang11prb}.
The nonzero Chern number corresponds to the $k_{z}$-dependent edge
states (known as the Fermi arc) according to the bulk-boundary correspondence
\cite{Hatsugai93prl}. Thus the two-node model in
Eq. (\ref{eq:model}) provides a generic description for Weyl semimetals,
including the band touching, opposite chirality, monopoles of Berry
curvature, topological charges, and Fermi arc. In the following we
shall focus on the topological case of $M_{0}M_{1}>0$.

\section{Landau bands}

\label{sec:Landau}
\begin{figure}[tbph]
\centering \includegraphics[width=0.48\textwidth]{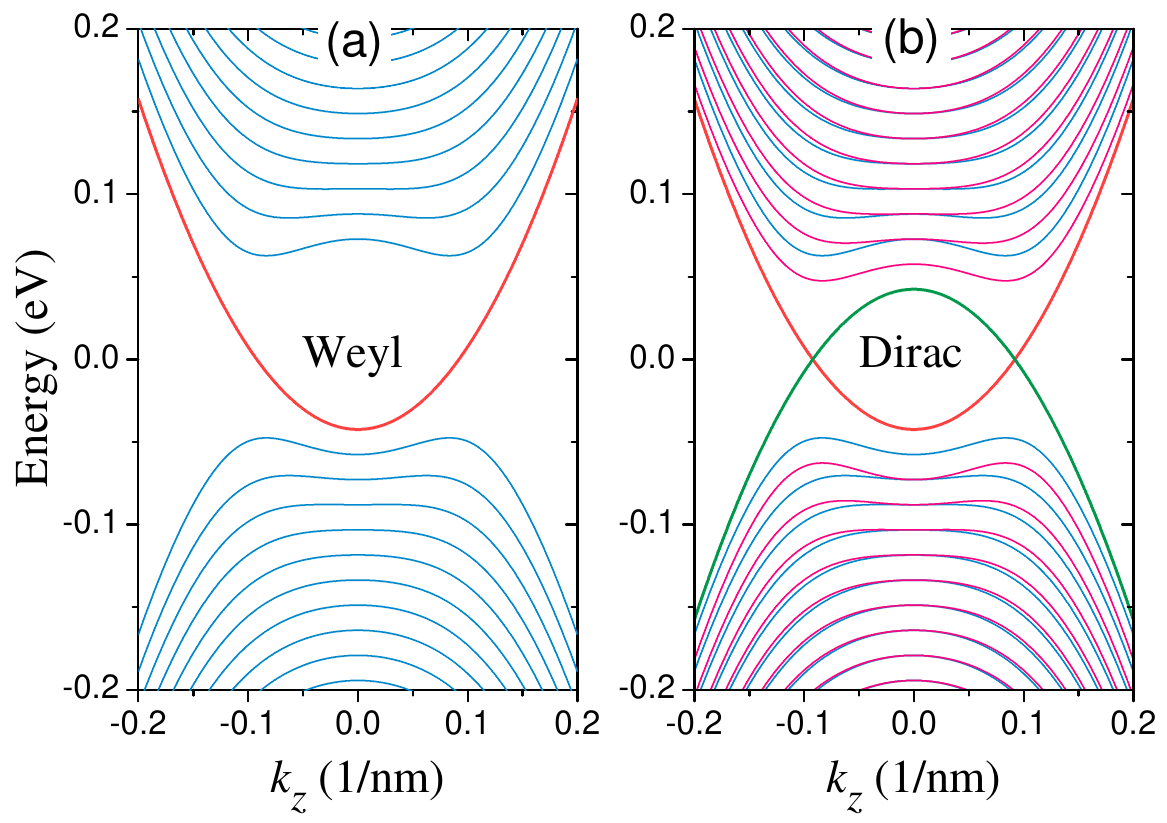}
\caption{The energies of Landau bands of the minimal global model for Weyl
and Dirac semimetals in a magnetic field $B$ applied along the $z$
direction, as functions of the wave vector $k_{z}$. The parameters:
$M_{0}=0.05$ eV, $M_{1}=5$ eV$\cdot$nm$^{2}$, $A=1$ eV$\cdot$nm,
and $B=1$ Tesla. The Zeeman energy is not included.}
\label{fig:Landau}
\end{figure}

In a magnetic field along the $z$ direction, the energy spectrum
is quantized into a set of 1D Landau bands dispersing with $k_{z}$
[see Fig.~\ref{fig:Landau} (a)]. We consider a magnetic field
applied along the $z$ direction, $\mathbf{B}=(0,0,B)$, and choose
the Landau gauge in which the vector potential is $\mathbf{A}=(-yB,0,0)$.
Under the Pierls replacement, the wave vector in the Hamiltonian in
Eq. (\ref{eq:model}) is replaced by the operator
\begin{equation}
\mathbf{k}=(k_{x}-\frac{eB_{z}}{\hbar}y,-i\partial_{y},k_{z})
\end{equation}
 $k_{x}$ and $k_{z}$ are still the good quantum numbers as the introduction
of the gauge field does not break the translational symmetry along
the x and z direction. Introducing the ladder operators \cite{Shen04prl,Shen04prb},
$k_{x}^{2}+k_{y}^{2}\rightarrow\omega(a^{\dag}a+1/2)$, $k_{+}\rightarrow(\sqrt{2}/\ell_{B})a^{\dag}$,
$k_{-}\rightarrow(\sqrt{2}/\ell_{B})a$, where the magnetic length
$\ell_{B}=\sqrt{\hbar/eB}$ and the ladder operators $a\equiv-[(y-\ell_{B}^{2}k_{x})/\ell_{B}+\ell_{B}\partial_{y}]/\sqrt{2}$
and $a^{\dag}\equiv-[(y-\ell_{B}^{2}k_{x})/\ell_{B}-\ell_{B}\partial_{y}]/\sqrt{2}$
\cite{Sakurai-book,Shen04prl}, then we can write the Hamiltonian
in terms of the ladder operators,
\begin{eqnarray}
H(\mathbf{k})=\left[\begin{array}{cc}
M_{k} & Ak_{-}\\
Ak_{+} & -M_{k}
\end{array}\right]\rightarrow\left[\begin{array}{cc}
M_{a} & \eta a\\
\eta a^{\dag} & -M_{a}
\end{array}\right],
\end{eqnarray}
where $\mathcal{M}_{a}=M_{0}-M_{1}k_{z}^{2}-\omega(a^{\dag}a+1/2)$,
$\omega=2M_{1}/\ell_{B}^{2}$, and $\eta=\sqrt{2}A/\ell_{B}$. With
the trial wave functions $(c_{1}|\nu-1\rangle,c_{2}|\nu\rangle)^{T}$
for $\nu=1,2,...$ (later denoted as $\nu\ge1$) and $(0,|0\rangle)^{T}$
for $\nu=0$, where $\nu$ indexes the Hermite polynomials, the eigen
energies $E$ can be found from the secular equation
\begin{equation}
\det\left[\begin{array}{cc}
\mathcal{M}_{\nu}+\omega/2-E & \eta\sqrt{\nu}\\
\eta\sqrt{\nu} & -\mathcal{M}_{\nu}+\omega/2-E
\end{array}\right]=0
\end{equation}
for $\nu\geq1$, and $-\mathcal{M}_{\nu}+\omega/2-E=0$ for $\nu=0$,
where $\mathcal{M}_{\nu}=M_{0}-M_{1}k_{z}^{2}-\omega\nu$. The eigen
energies are
\begin{eqnarray}\label{Energy-LL}
E_{k_{z}}^{\nu\pm} & = & \omega/2\pm\sqrt{\mathcal{M}_{\nu}^{2}+\nu\eta^{2}},\ \nu\ge1\nonumber \\
E_{k_{z}}^{0} & = & \omega/2-M_{0}+M_{1}k_{z}^{2},\ \ \nu=0.
\end{eqnarray}
This represents a set of Landau energy bands ($\nu$ as band index)
dispersing with $k_{z}$, as shown in Fig. \ref{fig:Landau}. The
eigen states for $\nu\ge1$ are
\begin{eqnarray}\label{LL-nu}
|\nu\ge1,k_{x},k_{z},+\rangle & = & \left[\begin{array}{cc}
\cos\frac{\theta_{k_{z}}^{\nu}}{2}|\nu-1\rangle\\
\sin\frac{\theta_{k_{z}}^{\nu}}{2}|\nu\rangle
\end{array}\right]|k_{x},k_{z}\rangle,\nonumber \\
|\nu\ge1,k_{x},k_{z},-\rangle & = & \left[\begin{array}{cc}
\sin\frac{\theta_{k_{z}}^{\nu}}{2}|\nu-1\rangle\\
-\cos\frac{\theta_{k_{z}}^{\nu}}{2}|\nu\rangle
\end{array}\right]|k_{x},k_{z}\rangle,
\end{eqnarray}
and for $\nu=0$ is
\begin{equation}\label{LL-nu0}
|\nu=0,k_{x},k_{z}\rangle=\left[\begin{array}{cc}
0\\
|0\rangle
\end{array}\right]|k_{x},k_{z}\rangle,
\end{equation}
where $\cos\theta=\mathcal{M}_{\nu}/\sqrt{\mathcal{M}_{\nu}^{2}+\nu\eta^{2}}$,
and the wave functions $\psi_{\nu,k_{z},k_{x}}(\mathbf{r})=\langle\mathbf{r}|\nu,k_{x},k_{z}\rangle$
are found as
\begin{eqnarray}\label{psi-nukxkz}
\psi_{\nu,k_{z},k_{x}}(\mathbf{r}) & = & \frac{C_{\nu}}{\sqrt{L_{x}L_{z}\ell_{B}}}e^{ik_{z}z}e^{ik_{x}x}e^{-\frac{(y-y_{0})^{2}}{2\ell_{B}^{2}}}\mathcal{H}_{\nu}(\frac{y-y_{0}}{\ell_{B}}),\nonumber \\
\end{eqnarray}
where $C_{\nu}\equiv1/\sqrt{\nu!2^{\nu}\sqrt{\pi}}$, $L_{x}L_{z}$
is area of sample, the guiding center $y_{0}=k_{x}\ell_{B}^{2}$,
$\mathcal{H}_{\nu}$ are the Hermite polynomials. As the dispersions
are not explicit functions of $k_{x}$, the number of different $k_{x}$
represents the Landau degeneracy $N_{L}=1/2\pi\ell_{B}^{2}=eB/h$
in a unit area in the x-y plane.

\begin{figure}
\centering\includegraphics[width=0.48\textwidth]{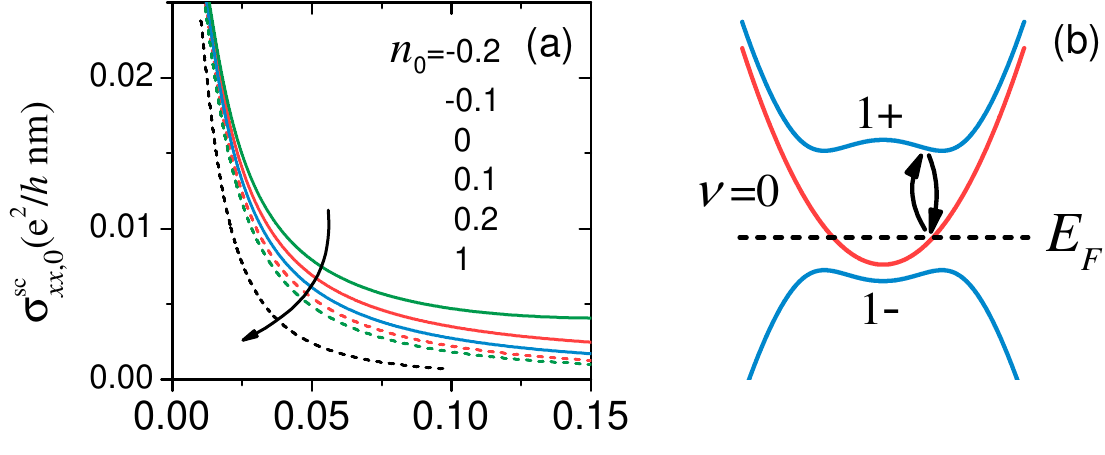}
\caption{(a) The $x$-direction conductivity of the 0-th Landau band of the
Weyl semimetal as a function of the $z$ direction magnetic field
$B$, for different values of the carrier density $n_{0}$ (in units
of $10^{17}$/cm$^{3}$). (b) Schematic of the second-order processes
that contribute to the $x$ direction conductivity. Parameters: $M_{0}=0.05$
eV, $M_{1}=5$ eV$\cdot$nm$^{2}$, $A=1$ eV$\cdot$nm, $V_{\text{imp}}=10$
(eV)$^{2}$$\cdot$nm$^{3}$.}
\label{Fig:sigma-SCx}
\end{figure}

This set of analytical solutions provides us a good base to study
the transport properties of Weyl fermions. In the following, we will
focus on the quantum limit, i.e., only the $\nu=0$ band is on the
Fermi surface [see Fig.~\ref{Fig:sigma-SCx}~(b)].

\section{z-direction semiclassical conductivity}

\label{sec:sigma-z}

\subsection{Argument of positive magnetoconductivity}

When the Fermi energy is located between the two states of $|\nu=1,k_{x},k_{z},\pm\rangle$,
all the bands for $|\nu\ge1,k_{x},k_{z},+\rangle$ are empty and all
the bands $|\nu\ge1,k_{x},k_{z},-\rangle$ are fully occupied. Only
the band of $\nu=0$ is partially filled. In this case the transport
properties of the system are dominantly determined by the highly degenerate
$\nu=0$ Landau bands [the red curve in Fig.~\ref{fig:Landau}
(a)] . It is reasonable to regard them as a bundle of one-dimensional
chains. Combining the Landau degeneracy $N_{L}$, the $z$-direction
conductance is approximately given by
\begin{eqnarray}
\sigma_{zz}=N_{L}\sigma_{\text{1D}},
\end{eqnarray}
where $\sigma_{\text{1D}}$ is the conductance for each one-dimensional
Landau band.

If we ignore the scattering between the states in the degenerate Landau
bands, according to the transport theory, the ballistic conductance
of a one-dimensional chain in the clean limit is given by
\begin{eqnarray}
\sigma_{\text{1D}}=\frac{e^{2}}{h},
\end{eqnarray}
then the conductivity is found as
\begin{eqnarray}\label{NMC-ballistic}
\sigma_{zz}=\frac{e^{2}}{h}\frac{eB}{h},
\end{eqnarray}
which is is linear in magnetic field $B$, giving a positive magnetoconductivity.

In most measurements, the sample size is much larger than the mean
free path, then the scattering between the states in the Landau bands
is inevitable, and we have to consider the other limit, i.e., the
diffusive limit. Usually, the scattering is characterized by a momentum
relaxation time $\tau$. According to the Einstein relation, the conductivity
of each Landau band in the diffusive limit is
\begin{eqnarray}
\sigma_{\text{1D}}=e^{2}N_{\text{1D}}v_{F}^{2}\tau,
\end{eqnarray}
where $v_{F}$ the Fermi velocity and the density of states for each
1D Landau band is $N_{\text{1D}}=1/\pi\hbar v_{F}$, then
\begin{eqnarray}\label{NMC-diffusive}
\sigma_{zz}=\frac{e^{2}}{h}\frac{eBv_{F}\tau}{\pi\hbar}.
\end{eqnarray}
If $v_{F}$ and $\tau$ are constant, one readily concludes that the
magnetoconductivity is positive and linear in $B$.

According to Nielsen and Ninomiya \cite{Nielsen83plb}, to illustrate the physical picture of the chiral anomaly, they started with a one-dimensional model in which two chiral energy bands have linear dispersions and opposite velocities.
An external electric field can accelerate electrons in one band to higher energy levels, in this way, charges are ``created". In contrast, in the other band, which has the opposite velocity, charges are annihilated. The chiral charge, defined as the difference between the charges in the two bands, therefore is not conserved in the electric field. This is literally the chiral anomaly. As one of the possible realizations of the one-dimensional chiral system, they then proposed to use the $\nu=0$ Landau bands of a three-dimensional semimetal, and expected ``the longitudinal magneto-conduction becomes extremely strong". In other words, the magnetoresistance of the 0th Landau bands in semimetals is the first physical quantity that was proposed as one of the signatures of the chiral anomaly.

Recently, several theoretical works have formulated the negative magnetoresistance or positive magnetoconductivity in the quantum limit as one of the signatures of the chiral anomaly \cite{Son13prb,Gorbar14prb}, much similar to those in Eqs. (\ref{NMC-ballistic}) and (\ref{NMC-diffusive}).
In both cases, the positive magnetoconductivity arises because the
Landau degeneracy increases linearly with $B$. However, in the following, we will show that if $v_{F}$ and $\tau$
also depend on the magnetic field, the conclusion has to be reexamined.

\subsection{Green function calculation}

Now we are ready to present the conductivity in the presence of the
magnetic field when the Fermi energy is located near the Weyl nodes.
The temperature is assumed to be much lower than the gap between bands
$1+$ and $1-$, i.e, $k_{B}T\ll2\sqrt{2}A/\ell_{B}$. In this case
all the Landau levels of $E_{k_{z}}^{\nu-}$ are fully occupied while
the $\nu=0$ band [the red curve in Fig.~\ref{fig:Landau} (a)]
is partially filled. Since $E_{k_{z}}^{0}$ is only a function of
$k_{z}$, and independent of $k_{x}$, the system can be regarded
as a bundle of highly degenerate one-dimensional chains. Along the
$z$ direction, the semiclassical Drude conductivity can be found
from the formula \cite{Datta1997}
\begin{eqnarray}
\sigma_{zz,0}^{sc}=\frac{e^{2}\hbar}{2\pi V}\sum_{k_{z},k_{x}}(v_{0}^{z})^{2}G_{0}^{R}G_{0}^{A},
\end{eqnarray}
where $-e$ is the electron charge, $V=L_{x}L_{y}L_{z}$ is the volume
with $L_{x}$ the length along the $x$ direction and so on, $v_{0}^{z}=\partial E_{k_{z}}^{0}/\hbar\partial k_{z}=2M_{1}k_{z}/\hbar$
is the velocity along the $z$ direction for a state with wave vector
$k_{z}$ in the $\nu=0$ band, $G_{0}^{R/A}=1/(E_{F}-E_{k_{z}}^{0}\pm i\hbar/2\tau^{0})$
is the retarded/advanced Green's function, with $\tau^{0}$ the lifetime
of a state in the $\nu=0$ band with wave vector $k_{x}$ and $k_{z}$.
Usually, in the diffusive regime, one can replace $G_{0k_{z}}^{R}G_{0k_{z}}^{A}$
by $\frac{2\pi}{\hbar}\tau_{k_{x}}^{0}\delta(E_{F}-E_{k_{z}}^{0})$.
However, in one dimension, to correct the van Hove singularity at
the band edge, we introduce an extra correction factor $\Lambda$,
so that $G_{0k_{z}}^{R}G_{0k_{z}}^{A}=\frac{2\pi}{\hbar}\Lambda\tau_{k_{x}}^{0}\delta(E_{F}-E_{k_{z}}^{0})$.
As shown in Appendix \ref{sec:Lambda}, $\Lambda\rightarrow1$ (0)
if the Fermi energy is far away from (approaching) the band edge.
Now the conductivity formula can be written as
\begin{eqnarray}
\sigma_{zz,0}^{sc} & = & \frac{e^{2}\hbar}{2\pi V}\sum_{k_{z},k_{x}}(v_{0k_{z}}^{z})^{2}\frac{2\pi}{\hbar}\Lambda\tau_{k_{x}}^{0}\delta(E_{F}-E_{k_{z}}^{0}).
\end{eqnarray}
The delta function $\delta(E_{F}-E_{k_{z}}^{0})=2\frac{\delta(k_{F}^{0}-k_{z})}{\hbar|v_{F}^{0}|}$,
where $v_{F}^{0}=2|M_{1}|k_{F}^{0}/\hbar$ is the absolute value of
the Fermi velocity of the $\nu=0$ band with $k_{F}^{0}$ the Fermi
wave vector. This allows us to perform the summation over $k_{z}$,
then
\begin{eqnarray}
\sigma_{zz,0}^{sc} & = & \frac{e^{2}}{h}\frac{2v_{F}^{0}}{L_{x}L_{y}}\sum_{k_{x}}\Lambda\tau_{k_{x}}^{0}.
\end{eqnarray}
The summation over $k_{x}$ is limited by the Landau degeneracy, finally
we can reduce the conductivity formula to
\begin{eqnarray}
\sigma_{zz,0}^{sc} & = & \frac{e^{2}}{h}\frac{v_{F}^{0}}{\pi L_{y}}\int_{-L_{y}/2\ell_{B}^{2}}^{L_{y}/2\ell_{B}^{2}}dk_{x}\tau^{0}\Lambda.\label{sigma-zz-0-tau}
\end{eqnarray}
The scattering time $\tau^{0}$ depends on the wave packet of the
Landau levels in band 0 and is a function of magnetic field. It can
be found from the iteration equation under the self-consistent Born
approximation (see Appendix \ref{sec:SCBA} for details)
\begin{eqnarray}
\hbar/2\tau^{0}=\sum_{k_{x}',k_{z}'}\frac{\langle|U_{k_{x}k_{z},k_{x}'k_{z}'}^{0,0}|^{2}\rangle(\hbar/2\tau^{0})}{[(E_{F}-E_{k_{z}}^{0})^{2}+(\hbar/2\tau^{0})^{2}]},
\end{eqnarray}
where $U_{k_{x}k_{z},k_{x}'k_{z}'}^{0,0}$ represents the scattering
matrix elements, $\langle...\rangle$ means the average over impurity
configurations. The conductivity in semimetals in vanishing magnetic
field has been discussed within the Born approximation \cite{DasSarma15prb}.

In this work, we consider only the short-range delta scattering potential.
The delta potential takes the form
\begin{eqnarray}
U(\mathbf{r})=\sum_{i}u_{i}\delta(\mathbf{r}-\mathbf{R}_{i}),
\end{eqnarray}
where $u_{i}$ measures the strength of scattering for an impurity
at $\mathbf{R}_{i}$, and the potential is delta correlated $\left\langle U\left(\mathbf{r}\right)U\left(\mathbf{r}'\right)\right\rangle =V_{\text{imp}}\delta(\mathbf{r}-\mathbf{r}')$,
where $V_{\text{imp}}$ is a field-independent parameter that is proportional
to the impurity density and averaged field-independent scattering
strength. Using the wave function of the $\nu=\text{0}$ band, we
find that (see Appendix \ref{sec:UU})
\begin{eqnarray}
\langle|U_{k_{x}k_{z},k_{x}'k_{z}'}^{0,0}|^{2}\rangle=\frac{V_{\text{imp}}\exp[-\ell_{B}^{2}(k_{x}-k_{x}')^{2}/2]}{L_{x}L_{z}\ell_{B}\sqrt{2\pi}},\label{UU-B-nu0}
\end{eqnarray}
and in the strong-field limit ($B\rightarrow\infty$),
\begin{eqnarray}
\tau^{0}\Lambda & = & \frac{\hbar^{2}v_{F}^{0}\pi\ell_{B}^{2}}{V_{\text{imp}}},\label{tau-00-strongB}
\end{eqnarray}
which gives the conductivity in the strong-field limit as
\begin{eqnarray}
\sigma_{zz,0}^{sc} & = & \frac{e^{2}}{h}\frac{(\hbar v_{F}^{0})^{2}}{V_{\text{imp}}}.\label{Eq:sigma-zz-0-v}
\end{eqnarray}
Notice that the Landau degeneracy in the scattering time cancels with
that in Eq. (\ref{sigma-zz-0-tau}), thus the magnetic field dependence
of $\sigma_{zz,0}^{sc}$ is given by the Fermi velocity $v_{F}^{0}$.
This is one of the main results in this paper. When ignoring the magnetic
field dependence of the Fermi velocity, a $B$-independent conductivity
was concluded, which is consistent with the previous work in which the velocity is constant \cite{Aji12prbr}.
Later, we will see the magnetic field dependence of the Fermi velocity can lead to different scenarios of positive and negative magnetoconductivity.

\section{Scenarios of negative and positive z-direction magnetoconductivity}

\label{sec:scenarios}

\subsection{Weyl semimetal with fixed carrier density}

\begin{figure}
\centering\includegraphics[width=0.48\textwidth]{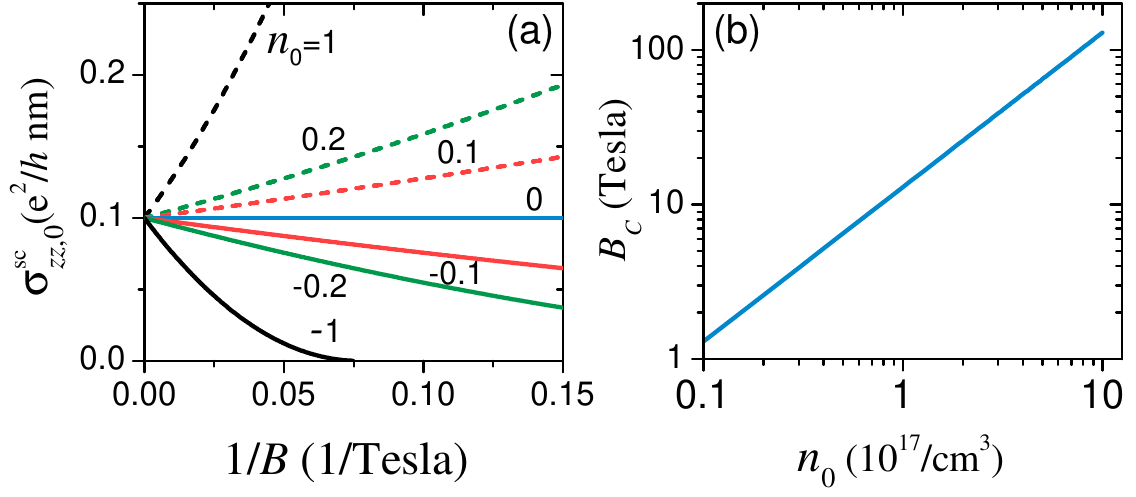}
\caption{(a) The $z$-direction conductivity of the 0-th Landau band of the
Weyl semimetal as a function of the $z$ direction magnetic field
$B$, for different values of the carrier density $n_{0}$ (in units
of $10^{17}$/cm$^{3}$). The lower bound of the field for each curve
is determined by the bottom of band $1+$ for $n_{0}>0$ or the bottom
of band 0 for $n_{0}<0$. (b) The characteristic $B_{c}$ defined
in Eq. (\ref{Eq:sigma-zz-0-B2}) as a function of $n_{0}$. Parameters:
$M_{0}=0.05$ eV, $M_{1}=5$ eV$\cdot$nm$^{2}$, $A=1$ eV$\cdot$nm,
$V_{\text{imp}}=10$ (eV)$^{2}$$\cdot$nm$^{3}$. Using $n_{0}=(8\pi/3)E_{F}^{3}/2M_{1}k_{c}A^{2}(2\pi)^{3}$,
a carrier density of $n_{0}\times10^{17}$/cm$^{3}$corresponds to
a Fermi energy of about $E_{F}=$144$\times{}^{3}\sqrt{n_{0}}$ meV.}
\label{Fig:sigma-SCz}
\end{figure}

In a strong field the Fermi velocity or the Fermi energy is given
by the density of charge carriers and the magnetic field \cite{Abrikosov98prb}.
We assume that an ideal Weyl semimetal is the case that the Fermi
energy crosses the Weyl nodes, all negative bands are fully filled
and the positive bands are empty. In this case $\hbar v_{F}^{0}=2M_{1}k_{c}$.
An extra doping of charge carriers will cause a change of electron
density $n_{0}(>0)$ in the electron-doped case or hole density $n_{0}(<0)$
in the hole-doped case. The relation between the Fermi wave vector
and the density of charge carriers is given by
\begin{equation}
n_{0}=2 N_{L}\times\frac{k_{F}^{0}-k_{c}}{2\pi}
\end{equation}
This means that  the  Fermi wave vector is determined by the density
of charge carriers $n_{0}$ and magnetic field $B$,
\begin{eqnarray}\label{kF-weyl-fixn0}
k_{F}^{0} & = & k_{c}+\pi n_{0}h/eB
\end{eqnarray}
or $k_{F}^{0}=k_{c}+2\pi^{2}n_{0}\ell_{B}^{2}$. Thus the Fermi velocity
is also a function of B, $\hbar v_{F}^{0}=2M_{1}k_{F}^{0},$ and
\begin{eqnarray}
\sigma_{zz,0}^{sc}=\sigma_{N}\left[1+\mathrm{sgn}(n_{0})\frac{B_{c}}{B}\right]^{2}.
\label{Eq:sigma-zz-0-B2}
\end{eqnarray}
where the characteristic field $B_{c}=\pi\left|n_{0}\right|h/ek_{c}$.
A typical order of $B_{c}$ is about 10 Tesla for $n_{0}$ of 10$^{17}$/cm$^{3}$
[see Fig.~\ref{Fig:sigma-SCz} (b)]. $\sigma_{zz,0}^{sc}$ is
constant for the undoped case of $n_{0}=0$, and
\begin{eqnarray}
\sigma_{N}=\frac{e^{2}}{h}\frac{4M_{1}^{2}k_{c}^{2}}{V_{\text{imp}}}\label{sigma-N}
\end{eqnarray}
is the conductivity of the undoped case, and is independent of magnetic
field. Thus the magnetoconductivity is always negative in the electron-doped
case while always positive in the hole-doped regime as shown in Fig.~\ref{Fig:sigma-SCz}
(a).

\subsection{Weyl semimetal with fixed Fermi energy}

In the case that the Fermi energy is fixed, $(\hbar v_{F}^{0})^{2}=4M_{1}(E_{F}-eM_{1}B/\hbar+M_{0})$,
and we have
\begin{eqnarray}
\sigma_{zz,0}^{sc} & = & \frac{e^{2}}{h}\frac{4M_{1}(E_{F}-eM_{1}B/\hbar+M_{0})}{V_{\text{imp}}},\label{Eq:MC-EF}
\end{eqnarray}
then the magnetoconductivity is always negative and linear in $B$.

\subsection{Dirac semimetal}

\label{sec:Dirac}

If the system has time-reversal symmetry, we may have a Dirac semimetal,
instead of Weyl semimetal, and all Weyl nodes turn to doubly-degenerate
Dirac nodes. A model for Dirac semimetal can be constructed \cite{Wang13prb}
by adding a time-reversal partner to Eq. (\ref{eq:model})
\begin{equation}
H_{\text{Dirac}}=\left[\begin{array}{cc}
H(\mathbf{k}) & 0\\
0 & H^{*}(-\mathbf{k})
\end{array}\right]+\sigma_{z}\otimes\left[\begin{array}{cc}
\Delta_{s} & 0\\
0 & \Delta_{p}
\end{array}\right].
\end{equation}
In the second term, the $z$-direction Zeeman energy $\Delta_{s/p}=g_{s/p}\mu_{B}B/2$
is also included, where $g_{s/p}$ is the g-factor for the $s/p$
orbital \cite{Jeon14natmat} and $\mu_{B}$ is the Bohr magneton.
Fig.~\ref{fig:Landau} (b) shows the Landau bands of both $H(\mathbf{k})$
and $H^{*}(-\mathbf{k})$ in the $z$-direction magnetic field. The
Landau bands of the Dirac semimetal can be found in a similar way
as that in Sec. \ref{sec:Landau}. Now there are two branches of $\nu=0$
bands, with the energy dispersions $E_{k_{z}}^{0\uparrow}=\omega/2+\Delta_{p}-M_{0}+M_{1}k_{z}^{2}$
and $E_{k_{z}}^{0\downarrow}=-\omega/2-\Delta_{s}+M_{0}-M_{1}k_{z}^{2}$
for $H(\mathbf{k})$ and $H^{*}(-\mathbf{k})$, respectively. They
intersect at $k_{z}=\pm\sqrt{[M_{0}-(\omega+\Delta_{s}+\Delta_{p})/2]/M_{1}}$
and energy $(\Delta_{p}-\Delta_{s})/2$, and with opposite Fermi velocities
near the points. In the absence of inter-block velocity, the longitudinal
conductance along the $z$ direction is approximately a summation
of those for two independent Weyl semimetals.

First, we consider the Fermi energy cross both bands $0\uparrow$
and $0\downarrow$. Using Eq. (\ref{Eq:MC-EF}), the $z$-direction
conductivity is found as
\begin{eqnarray}
\sigma_{zz,0}^{sc} & = & \sigma_{zz,0\uparrow}^{sc}+\sigma_{zz,0\downarrow}^{sc}\nonumber \\
 & = & \frac{e^{2}}{h}\frac{8M_{1}}{V_{\text{imp}}}[M_{0}-\frac{eM_{1}B}{\hbar}-\frac{\mu_{B}(g_{p}+g_{s})B}{4}],
\end{eqnarray}
or using $\sigma_{N}$ defined in Eq. (\ref{sigma-N}),
\begin{eqnarray}
\sigma_{zz,0}^{sc} & = & 2\sigma_{N}[1-\frac{eB}{\hbar k_{c}^{2}}-\frac{\mu_{B}(g_{p}+g_{s})B}{4M_{0}}].\label{Eq:MC-Dirac}
\end{eqnarray}
In this case we have a negative linear $B$ magnetoconductivity, when
the Fermi energy crosses both $E_{k_{z}}^{0\uparrow}$ and $E_{k_{z}}^{0\downarrow}$.

With increasing magnetic field, the $0\uparrow$ bands will shift
upwards and the $0\downarrow$ bands will shift downwards. Beyond
a critical field, the Fermi energy will fall into either $0\uparrow$
or $0\downarrow$ bands, depending on whether the carriers are electron-type
or hole-type. If the carrier density is fixed, the Fermi wave vector
in this case does not depend on $k_{c}$ as that in Eq. (\ref{kF-weyl-fixn0}),
but
\begin{eqnarray}
k_{F}^{0} & = & \frac{\pi n_{0}h}{eB}
\end{eqnarray}
or $k_{F}^{0}=2\pi^{2}n_{0}\ell_{B}^{2}$. In this case, with increasing
magnetic field, the Fermi energy will approach the band edge and the
Fermi velocity always decreases. Using Eq. (\ref{Eq:sigma-zz-0-v}),
\begin{eqnarray}
\sigma_{zz,0}^{sc} & = & \frac{e^{2}}{h}\frac{4\pi^{2}h^{2}M_{1}^{2}n_{0}^{2}}{V_{\text{imp}}e^{2}B^{2}},\label{sigma-zz-dirac-largeB}
\end{eqnarray}
which also gives negative magnetoconductivity that is independent
on the type of carriers.

Note that in the Weyl semimetal TaAs with broken inversion symmetry,
where the Weyl nodes always come in even pairs because of time-reversal
symmetry \cite{Weng15prx,Huang15arXiv,Lv15arXiv,Xu15arXiv}, the situation
is more similar to that for the Dirac semimetal and the magnetoconductivity
does not depend on the type of carriers and may be described by a
generalized version of Eqs. (\ref{Eq:MC-Dirac}) and (\ref{sigma-zz-dirac-largeB}).

\section{Longitudinal and Hall conductivities in x-y plane}

\label{sec:sigma-x}

\subsection{$x$-direction conductivity}

In the $x-y$ plane normal to the magnetic field, the longitudinal
conductivity along either the $x$ or $y$ direction is negligibly
small as the effective velocity 
\begin{eqnarray}
v_{0}^{x}=\frac{\partial E_{k_{z}}^{0}}{\hbar\partial k_{x}}=0.
\end{eqnarray}
Nevertheless, a non-zero longitudinal conductivity along the $x$ direction can
be found as 
\begin{eqnarray}
\sigma_{xx,0}^{sc}=\frac{e^{2}\hbar}{\pi V}\sum_{k_{x},k_{z},\nu=\pm}\mathrm{Re}(G_{0}^{R}v_{0,1\nu}^{x}G_{1\nu}^{A}v_{1\nu,0}^{x}),
\end{eqnarray}
where the Green's functions of band $1\pm$ are $G_{1\pm}^{R/A}=1/(E_{F}-E_{k_{z}}^{1\pm}\pm i\hbar/2\tau_{k_{x},k_{z}}^{1\pm})$,
and the inter-band velocity
\begin{eqnarray}\label{v-x-0-u+}
\hbar v_{0,1+}^{x}&=&\frac{\sqrt{2}M_{1}}{\ell_{B}}\sin \frac{\theta_{k_{z}}^{1}}{2}
+ A\cos\frac{\theta_{k_{z}}^{1}}{2}\nonumber\\
\hbar v_{0,1-}^{x}&=&- \frac{\sqrt{2}M_{1}}{\ell_{B}}\cos\frac{\theta_{k_{z}}^{1}}{2}
+A\sin\frac{\theta_{k_{z}}^{1}}{2}.
\end{eqnarray}
Note that for the Landau bands generated
by the $z$-direction magnetic field, the leading-order $x$-direction
velocity is the inter-band velocity $v_{0,1\pm}^{x}$ that couples
band $0$ with bands $1\pm$, and $\tau_{k_{x}k_{z}}^{1\pm\leftrightarrow0}$,
the scattering times of band $1\pm$ are due to virtual scattering
going back and forth between band 0 [see Fig.~\ref{Fig:sigma-SCz}
(d)], so $\sigma_{xx,0}^{sc}$ indeed stems from second-order processes
and therefore is much smaller than $\sigma_{zz,0}^{sc}$ that arises
from first-order processes. We find that $\sigma_{xx,0}^{sc}=\sigma_{xx}^{0,1+}+\sigma_{xx}^{0,1-}$,
where (see Appendix \ref{sigma-01-app} for details)
\begin{eqnarray}
\sigma_{xx}^{0,1\pm}=\frac{e^{2}\hbar}{V}\sum_{k_{x},k_{z}}|v_{0,1\pm}^{x}|^{2}\frac{\Lambda\delta(E_{F}-E_{k_{z}}^{0})}{2(E_{F}-E_{k_{z}}^{1\pm})^{2}}\frac{\hbar}{\tau_{k_{x}k_{z}}^{1\pm\leftrightarrow0}}.\label{Eq:sigma-xx-01-1-1}
\end{eqnarray}
At this stage, we have the same form as Eq. (28) in the paper by Abrikosov
\cite{Abrikosov98prb}. If the Hamiltonian is replaced by $H=vk\cdot\sigma$
and the scattering time is evaluated for screened charge impurities
under the random phase approximation, an $x$-direction $1/B$ magnetoconductivity
can be found, leading to the quantum linear magnetoresistance. In
the present case, the scattering time is found as
\begin{eqnarray}
\frac{\hbar}{\tau_{k_{z}k_{x}}^{1\pm\leftrightarrow0}}=\frac{V_{\text{imp}}}{2\pi\ell_{B}^{2}}\frac{\Lambda}{\hbar v_{F}^{0}}(1\mp\cos\theta_{k_{z}=k_{F}^{0}}^{1}).\label{tau-01}
\end{eqnarray}
Note that $\sigma_{zz,0}^{sc}$ is proportional to $\tau^{0}$ so
the Landau degeneracy $1/2\pi\ell_{B}^{2}$ from the conductivity
formula cancels with that from the scattering time. However, $\sigma_{xx,0}^{sc}$
is inversely proportional to the scattering time $\tau_{k_{x}k_{z}}^{1+}$
then the effect of the Landau degeneracy actually is doubled, and
finally we arrive at
\begin{eqnarray}\label{sigma-xx-01+}
\sigma_{xx}^{0,1+}=\frac{e^{2}}{h}\frac{V_{\text{imp}}}{2\pi^{2}\ell_{B}^{4}}\left[\frac{\Lambda^{2}}{(v_{F}^{0})^{2}}\frac{|v_{0,1+}^{x}(k_{z})|^{2}}{(E_{F}-E_{k_{z}}^{1+})^{2}}\sin^{2}\frac{\theta_{k_{z}}^{1}}{2}\right]_{k_{z}=k_{F}^{0}},\nonumber \\
\end{eqnarray}
where one replaces $\sin$ by $\cos$ for $\sigma_{xx}^{0,1-}$. In
both the electron- and hole-doped regimes, the magnetoconductivity
is always positive as shown in Fig.~\ref{Fig:sigma-SCz} (c).

In the strong-field limit, $\sin^{2}(\theta_{k_{z}}^{1}/2)\rightarrow1$
if $M_{1}>0$, $\hbar\hat{v}_{0,\mu+}^{x}\rightarrow M_{1}\sqrt{2}/\ell_{B}$
according to Eq. (\ref{v-x-0-u+}), and $(E_{F}-E_{k_{z}}^{1+})^{2}\rightarrow\omega^{2}=4M_{1}^{2}/\ell_{B}^{4}$
according to Eq. (\ref{Energy-LL}), then
\begin{eqnarray}
\sigma_{xx}^{0,1}=\frac{e^{2}}{h}\frac{V_{\text{imp}}}{4\pi^{2}\ell_{B}^{2}}(\frac{\Lambda}{ \hbar v_{F}^{0}})^2.\label{sigma-xx-01+-strongB}
\end{eqnarray}

(1) In Weyl semimetals with a fixed carrier density, the magnetic
field will push the Fermi wave vector to $k_{F}^{0}=k_{c}$, near
which $\Lambda\rightarrow1$ and $\hbar v_{F}^{0}=2M_{1}k_{c}(1+\text{sgn}(n_{0})B_{c}/B)$,
then
\begin{eqnarray}
\sigma_{xx}^{0,1}=\frac{e^{2}}{h}\frac{eV_{\text{imp}}B}{16\pi^{2}hM_{1}^{2}k_{c}^{2}(1+\text{sgn}(n_{0})B_{c}/B)^{2}}.
\end{eqnarray}
In the limit that $B\gg B_{c}$, $\sigma_{xx}^{0,1}$ increases linearly
with $B$.

(2) In Dirac semimetals, the magnetic field pushes the Fermi energy
to the band edge, using $\Lambda/v_F^0$ in Eq. (\ref{Lambda-vF-bandbottom}),
\begin{eqnarray}
\sigma_{xx}^{0,1}=\frac{e^{2}}{h}\frac{V_{\text{imp}}}{4\pi^{2}\ell_{B}^{2}}\left(\frac{\pi\ell_{B}^{2}}{4M_{1}V_{\text{imp}}}\right)^{2/3}\propto B^{1/3}.\label{sigma-xx-01+-dirac-strongB}
\end{eqnarray}

\subsection{Hall conductivity}

In the presence of the $z$-direction magnetic field, a Hall conductance
in the $x-y$ plane can also be generated \cite{Yang11prb,Zyuzin12prb,Goswami13prb,Burkov14prl}.
The correction of an electric field $E_{y}$ to the model Hamiltonian
is the potential energy,
\begin{equation}
\Delta V=-eE_{y}y.
\end{equation}
In the state of $\nu=0$, the energy dispersion is corrected to $E_{k_{z}}^{0}-eE_{y}\ell_{B}^{2}k_{x}$ as $\left\langle y\right\rangle =l_{B}^{2}k_{x}$.
This energy correction leads to a velocity shift along the $x$ direction,
\begin{eqnarray}
v_{x}\equiv\frac{1}{\hbar}\frac{\partial(E_{k_{z}}^{0}-eE_{y}\ell_{B}^{2}k_{x})}{\partial k_{x}}=-\frac{eE_{y}\ell_{B}^{2}}{\hbar}.
\end{eqnarray}
For each $k_{z}$, this $v_{x}$ leads to a quantized Hall conductance
\begin{eqnarray}
\frac{j_{x}}{E_{y}}=-\frac{ev_{x}}{E_{y}}\times\frac{1}{2\pi\ell_{B}^{2}}=\frac{e^{2}}{h}.
\end{eqnarray}
The total Hall conductance is found by integrating over $k_{z}$ up
to the Fermi wave vector $k_{F}^{0}$, and
\begin{eqnarray}
\sigma_{yx}=k_{F}^{0}\frac{e^{2}}{\pi h}.
\end{eqnarray}
In particular, $k_{F}^{0}=k_{c}+\pi n_{0}h/eB$ for Weyl semimetals with a fixed carrier
density $n_{0}$, and a Hall conductance is found as
\begin{equation}
\sigma_{yx}=\frac{en_{0}}{B}+\frac{e^{2}k_{c}}{h\pi}
\end{equation}
The first term is attributed to  the classical Hall effects, and
the second term comes from the non-zero Chern number of the fully
filled low energy bands of $-k_{c}<k_{z}<k_{c}$. This is consistent
with the calculation by using the Kubo formula for the Hall conductivity
\cite{ZhangSB14prb}.

\section{Summary}

\label{sec:summary}

We present a conductivity formula for the lowest Landau band in a
semimetal in the presence of short-range delta scattering potentials.
The conductivity depends on the square of the Fermi velocity, instead
of the Landau degeneracy. Based on this mechanism and the model that
describes two Weyl nodes with a finite spacing in momentum space,
we find three cases that could give a negative magnetoconductivity
in the strong-field limit. (i) A Weyl semimetal with a fixed density
of electron-type charge carries [Eq. (\ref{Eq:sigma-zz-0-B2})]. (ii)
A Weyl semimetal with a fixed Fermi energy [Eq. (\ref{Eq:MC-EF})].
(iii) A Dirac semimetal or a Weyl semimetal with time-reversed pairs
of Weyl nodes [Eqs. (\ref{Eq:MC-Dirac}) and (\ref{sigma-zz-dirac-largeB})], with a $1/B^2$ dependence.
These formulas are valid as long as the Fermi energy crosses the $\nu=0$
Landau bands. Our theory can be applied to account for the negative
magnetoconductivity observed experimentally in various topological
semimetals in high magnetic fields, such as BiSb alloy \cite{Kim13prl},
TaAs \cite{HuangXC15arXiv,ZhangCL15arXiv}, and Na$_{3}$Bi \cite{Xiong15arXiv}. In this
way, we conclude that a positive magnetoconductivity (or negative
magnetoresistance) in the strong-field limit is not a compelling signature
of the chiral anomaly in topological semimetals. Our theory can also
be generalized to understand the magnetoconductivity in non-topological
three-dimensional semimetals.

\begin{acknowledgements} This work was supported by the Research
Grant Council, University Grants Committee, Hong Kong under Grant
No.: 17303714, and HKUST3/CRF/13G. \end{acknowledgements}

\appendix

\section{About the correction factor $\Lambda$}

\label{sec:Lambda}

For the $\nu=0$ band, we need to deal with the imaginary part of
the Green's function
\begin{eqnarray}
\frac{\frac{\hbar}{2\tau_{k_{x}}^{0}}}{[M_{1}k_{z}^{2}-(E_{F}+M_{0}-\omega/2)]^{2}+(\frac{\hbar}{2\tau_{k_{x}}^{0}})^{2}},
\end{eqnarray}
In this work, we assume $M_{0},M_{1},E_{F},\omega>0$. In this case,
we can write $a=\sqrt{M_{1}}$, $b=\sqrt{E_{F}+M_{0}-\omega/2}=\hbar v_{F}^{0}/2\sqrt{M_{1}}$,
and $c=2\tau_{k_{x}}^{0}/\hbar$. A widely used approximation is that
\begin{eqnarray}
\int_{-\infty}^{\infty}dx\frac{1/c}{(a^{2}x^{2}-b^{2})^{2}+1/c^{2}}=\int_{-\infty}^{\infty}dx\pi\delta(a^{2}x^{2}-b^{2}).\nonumber \\
\end{eqnarray}
However, this leads to unphysical van Hove singularities at the band
edges. We correct this approximation with an extra factor $\Lambda$,
so that
\begin{eqnarray}
\int_{-\infty}^{\infty}dx\frac{1/c}{(a^{2}x^{2}-b^{2})^{2}+1/c^{2}}=\int_{-\infty}^{\infty}dx\pi\Lambda\delta(a^{2}x^{2}-b^{2}).\nonumber \\
\end{eqnarray}
The form of $\Lambda$ can be found as follows. First, the integral
can be found as
\begin{eqnarray}
\int_{-\infty}^{\infty}dx\frac{1/c}{(a^{2}x^{2}-b^{2})^{2}+1/c^{2}}=\frac{\sqrt{c}}{a}\frac{\pi}{\sqrt{2}}\frac{\sqrt{C^{2}+\sqrt{C^{4}+1}}}{\sqrt{C^{4}+1}},\nonumber \\
\end{eqnarray}
where $C^{2}=b^{2}c$. On the other hand, using the property of the
delta function
\begin{eqnarray}
\pi\Lambda\int_{-\infty}^{\infty}dx\delta(a^{2}x^{2}-b^{2})=\pi\Lambda/ab.\label{f-integral}
\end{eqnarray}
So
\begin{eqnarray}
\Lambda=\frac{\sqrt{C^{4}+C^{2}\sqrt{C^{4}+1}}}{\sqrt{2(C^{4}+1)}}.\label{Eq:Lambda}
\end{eqnarray}
In the limit $C\gg1$, $\Lambda\rightarrow1$, while in the other
limit $C\ll1$,
\begin{eqnarray}
\lim_{C\ll1}\Lambda^{2}=\frac{(\hbar v_{F}^{0})^{2}}{4M_{1}}\frac{\tau^{0}}{\hbar},\label{Lambda-bandbottom}
\end{eqnarray}
so this correction is necessary near the band bottom, where $E_{F}+M_{0}-\omega/2=0$,
and can be ignored when the Fermi energy is far away from the band
bottom.

Now we evaluate $\Lambda/\hbar v_{F}^{0}$ in the band bottom limit.
Combining Eqs. (\ref{tau-00-strongB}) and (\ref{Lambda-bandbottom}),
we arrive at
\begin{eqnarray}
\lim_{C\ll1}\frac{\Lambda}{\hbar v_{F}^{0}}=\left(\frac{\pi\ell_{B}^{2}}{4M_{1}V_{\text{imp}}}\right)^{1/3}\propto B^{-1/3}.\label{Lambda-vF-bandbottom}
\end{eqnarray}

\section{Self-consistent Born approximation}

\label{sec:SCBA} In the self-consistent Born approximation, the full
Green function is written as
\begin{eqnarray}
G_{0k_{z}k_{x}}^{R}(E_{F}) & = & \frac{1}{E_{F}-E_{k_{z}k_{x}}^{0}-\Sigma_{0k_{z}k_{x}}^{R}(E_{F})},
\end{eqnarray}
where the self-energy
\begin{eqnarray}
\Sigma_{0k_{z}k_{x}}^{R}(E_{F}) & = & \sum_{k_{x}',k_{z}'}\langle|U_{k_{z}k_{x},k_{z}'k_{x}'}^{0,0}|^{2}\rangle G_{0k_{z}'k_{x}'}^{R}(E_{F}).
\end{eqnarray}
The real part of the self-energy can be absorbed into the definition
of the chemical potential, we only need the imaginary part,
\begin{eqnarray}
\mathcal{I} & = & \frac{V_{\text{imp}}}{L_{x}L_{z}\ell_{B}\sqrt{2\pi}}\sum_{k_{x}',k_{z}'}\frac{\mathcal{I}e^{-\frac{\ell_{B}^{2}(k_{x}-k_{x}')^{2}}{2}}}{(E_{F}-E_{k_{z}'}^{0})^{2}+\mathcal{I}^{2}}.
\end{eqnarray}
where we have used Eq. (\ref{UU-B-nu0}) and suppressed the $k_{x}$
and $k_{z}$ dependence of $\mathcal{I}$ in the strong-field limit.
Using $\frac{1}{L_{z}}\sum_{k_{z}'}=\int_{-\infty}^{\infty}\frac{dk_{z}'}{2\pi}$,
$\frac{1}{L_{x}}\sum_{k_{x}'}=\int_{-L_{y}/2\ell_{B}^{2}}^{L_{y}/2\ell_{B}^{2}}\frac{dk_{x}'}{2\pi}=\frac{1}{\ell_{B}}\int_{\ell_{B}k_{x}-L_{y}/2\ell_{B}}^{\ell_{B}k_{x}+L_{y}/2\ell_{B}}\frac{dK}{2\pi}$,
where $K=\ell_{B}(k_{x}-k_{x}')$, and in the strong-field limit,
$\ell_{B}k_{x}\pm L_{y}/2\ell_{B}\rightarrow\pm\infty$,
\begin{eqnarray}
\mathcal{I} & = & \frac{V_{\text{imp}}}{2\pi\ell_{B}^{2}}\int_{-\infty}^{\infty}\frac{dk_{z}'}{2\pi}\frac{\mathcal{I}}{(E_{F}-E_{k_{z}'}^{0})^{2}+\mathcal{I}^{2}}.
\end{eqnarray}
Using $E_{k_{z}}^{0}=\omega/2-M_{0}+M_{1}k_{z}^{2}$,
\begin{eqnarray}
\mathcal{I} & = & \frac{V_{\text{imp}}}{2\pi\ell_{B}^{2}}\int_{-\infty}^{\infty}\frac{dk_{z}'}{2\pi}\frac{\mathcal{I}}{[M_{1}(k_{z}')^{2}-E_{F}+\omega/2-M_{0}]^{2}+\mathcal{I}^{2}}.\nonumber \\
\end{eqnarray}
We consider $E_{F}+M_{0}>\omega/2$, so the integral can be written
into
\begin{eqnarray}
\frac{1}{c}=\frac{V_{\text{imp}}}{(2\pi)^{2}\ell_{B}^{2}}\int_{-\infty}^{\infty}dx\frac{1/c}{(a^{2}x^{2}-b^{2})^{2}+1/c^{2}},
\end{eqnarray}
with $a=\sqrt{M_{1}}$, $b=\sqrt{E_{F}+M_{0}-\omega/2}$, $c=1/\mathcal{I}$.
Using Eq. (\ref{f-integral}),
\begin{eqnarray}
\int_{-\infty}^{\infty}dx\frac{1/c}{(a^{2}x^{2}-b^{2})^{2}+1/c^{2}}=\frac{\pi}{ab}\Lambda,
\end{eqnarray}
then
\begin{eqnarray}
\frac{\hbar}{2\tau^{0}}=\frac{1}{c}=\frac{V_{\text{imp}}}{(2\pi)^{2}\ell_{B}^{2}}\frac{\pi}{ab}\Lambda=\frac{V_{\text{imp}}}{2\pi\ell_{B}^{2}}\frac{\Lambda}{\hbar v_{F}^{0}}.
\end{eqnarray}

\section{Scattering matrix elements}

\label{sec:UU}

We calculate a general form of the scattering matrix element $\langle U_{k_{x}'k_{z}',k_{x}k_{z}}^{\alpha,\beta}U_{k_{x}k_{z},k_{x}'k_{z}'}^{\gamma,\delta}\rangle$,
where
\begin{eqnarray}
U_{k_{x}'k_{z}',k_{x}k_{z}}^{\alpha,\beta} & \equiv & \int d\mathbf{r}'\langle\alpha,k_{x}',k_{z}'|\mathbf{r}'\rangle U(\mathbf{r}')\langle\mathbf{r}'|\beta,k_{x},k_{z}\rangle,\nonumber \\
U_{k_{x}k_{z},k_{x}'k_{z}'}^{\gamma,\delta} & \equiv & \int d\mathbf{r}\langle\gamma,k_{x},k_{z}|\mathbf{r}\rangle U(\mathbf{r})\langle\mathbf{r}|\delta,k_{x}',k_{z}'\rangle.
\end{eqnarray}
Using the wave functions in Eq. (\ref{psi-nukxkz}) and the identity
\begin{eqnarray}
 &  & \langle\int d\mathbf{r}\int d\mathbf{r}'f(\mathbf{r})f(\mathbf{r}')U(\mathbf{r})U(\mathbf{r}')\rangle\nonumber \\
 & = & \int d\mathbf{r}\int d\mathbf{r}'f(\mathbf{r})f(\mathbf{r}')\langle U(\mathbf{r})U(\mathbf{r}')\rangle,
\end{eqnarray}
and $\langle U(\mathbf{r})U(\mathbf{r}')\rangle=V_{\mathrm{imp}}\delta(\mathbf{r}-\mathbf{r}')$,
we have
\begin{eqnarray}
 &  & \langle U_{k_{x}'k_{z}',k_{x}k_{z}}^{\alpha,\beta}U_{k_{x}k_{z},k_{x}'k_{z}'}^{\gamma,\delta}\rangle\nonumber \\
 & = & \frac{V_{\text{imp}}C_{\alpha}C_{\beta}C_{\gamma}C_{\delta}}{L_{x}^{2}L_{z}^{2}\ell_{B}^{2}}\int d\mathbf{r}e^{-\frac{(y-y_{0}')^{2}}{\ell_{B}^{2}}}e^{-\frac{(y-y_{0})^{2}}{\ell_{B}^{2}}}\nonumber \\
 &  & \times\mathcal{H}_{\alpha}(\frac{y-y_{0}'}{\ell_{B}})\mathcal{H}_{\beta}(\frac{y-y_{0}}{\ell_{B}})\mathcal{H}_{\gamma}(\frac{y-y_{0}}{\ell_{B}})\mathcal{H}_{\delta}(\frac{y-y_{0}'}{\ell_{B}}).\nonumber \\
\end{eqnarray}
After performing the integral, we have
\begin{eqnarray}
\langle U_{k_{x}'k_{z}',k_{x}k_{z}}^{\alpha,\beta}U_{k_{x}k_{z},k_{x}'k_{z}'}^{\gamma,\delta}\rangle & = & \frac{V_{\text{imp}}}{L_{x}L_{z}\ell_{B}}I_{K}^{\alpha\beta\gamma\delta},\label{U-abcd-integral}
\end{eqnarray}
where we have defined a dimensionless integral
\begin{eqnarray}\label{IH-K}
I_{K}^{\alpha\beta\gamma\delta} & = & C_{\alpha}C_{\beta}C_{\gamma}C_{\delta}e^{-\frac{K^{2}}{2}}\int_{-\infty}^{\infty}dXe^{-2X^{2}}\mathcal{H}_{\alpha}(X+\frac{K}{2})\nonumber \\ &  & \times\mathcal{H}_{\beta}(X-\frac{K}{2})\mathcal{H}_{\gamma}(X-\frac{K}{2})\mathcal{H}_{\delta}(X+\frac{K}{2}),
\end{eqnarray}
and $K=(y_{0}-y_{0}')/\ell_{B}$. Using Eqs. (\ref{U-abcd-integral})
and (\ref{IH-K}), it is straightforward to find Eq. (\ref{UU-B-nu0}).

\section{Calculation of $x$-direction conductivity }

\label{sigma-01-app}

Now we evaluate
\begin{eqnarray}
\sigma_{xx}^{0,1\pm} & = & \frac{e^{2}\hbar}{2\pi V}\sum_{k_{x},k_{z}}2\text{Re}\left(G_{0}^{R}v_{0,1\pm}^{x}G_{1\pm}^{A}v_{1\pm,0}^{x}\right),
\end{eqnarray}
where the Green's functions
\begin{eqnarray}
G_{0}^{R} & = & \frac{1}{E_{F}-E_{k_{z}}^{0}+i\frac{\hbar}{2\tau_{k_{x}}^{0}}},\ \ \ G_{1\pm}^{A}=\frac{1}{E_{F}-E_{k_{z}}^{1\pm}-i\frac{\hbar}{2\tau_{k_{x}k_{z}}^{1\pm}}},\nonumber \\
\end{eqnarray}
and the velocities along $x$ direction are found as
\begin{eqnarray}
\hbar\hat{v}_{0,\mu+}^{x} & = & (\frac{M_{1}\sqrt{2}}{\ell_{B}}\sin\frac{\theta_{k_{z}}^{\mu}}{2}+A\cos\frac{\theta_{k_{z}}^{\mu}}{2})\delta_{\mu,1},\nonumber \\
\hbar\hat{v}_{0,\mu-}^{x} & = & (-\frac{M_{1}\sqrt{2}}{\ell_{B}}\cos\frac{\theta_{k_{z}}^{\mu}}{2}+A\sin\frac{\theta_{k_{z}}^{\mu}}{2})\delta_{\mu,1}.
\end{eqnarray}
then
\begin{eqnarray}
\sigma_{xx}^{0,1+} & \approx & \frac{e^{2}\hbar}{2\pi V}\sum_{k_{x},k_{z}}2|v_{0,1+}^{x}|^{2}\frac{\frac{\hbar}{2\tau_{k_{x}}^{0}}}{[(E_{F}-E_{k_{z}}^{0})^{2}+(\frac{\hbar}{2\tau_{k_{x}}^{0}})^{2}]}\nonumber \\
 &  & \times\frac{\frac{\hbar}{2\tau_{k_{x}k_{z}}^{1+}}}{[(E_{F}-E_{k_{z}}^{1+})^{2}+(\frac{\hbar}{2\tau_{k_{x}k_{z}}^{1+}})^{2}]}.
\end{eqnarray}
We assume that the Fermi energy crosses only the $\nu=0$ Landau band,
in this case,
\begin{eqnarray}
\frac{\frac{\hbar}{2\tau_{k_{x}}^{0}}}{[(E_{F}-E_{k_{z}}^{0})^{2}+(\frac{\hbar}{2\tau_{k_{x}}^{0}})^{2}]}\approx\pi\Lambda\delta(E_{F}-E_{k_{z}}^{0}).
\end{eqnarray}
Note that different from Abrikosov's approximation in Eqs. (27)-(28)
of Ref. \onlinecite{Abrikosov98prb}, an extra correction factor
$\Lambda$ is added. Then
\begin{eqnarray}
\sigma_{xx}^{0,1+} & = & \frac{e^{2}\hbar}{V}\sum_{k_{x},k_{z}}|v_{0,1+}^{x}|^{2}\frac{\Lambda\delta(E_{F}-E_{k_{z}}^{0})\frac{\hbar}{2\tau_{k_{x}k_{z}}^{1+}}}{(E_{F}-E_{k_{z}}^{1+})^{2}+(\frac{\hbar}{2\tau_{k_{x}k_{z}}^{1+}})^{2}}.\nonumber \\
\end{eqnarray}
Using $\frac{1}{L_{z}}\sum_{k_{z}}=\int\frac{dk_{z}}{2\pi}$, $\frac{1}{L_{x}}\sum_{k_{x}}=\int_{-L_{y}/2\ell_{B}^{2}}^{L_{y}/2\ell_{B}^{2}}\frac{dk_{x}}{2\pi}$,
$\delta(E_{F}-E_{k_{z}}^{0})=\sum_{i}\frac{\delta(k_{F}^{0,i}-k_{z})}{\hbar|v_{0k_{z}}^{z}|}$,
where $k_{z}=k_{F}^{0,i}$ are the roots of $E_{F}=E_{k_{z}}^{0}$,
and for the $\nu=0$ band, $k_{F}^{0,+}=-k_{F}^{0,-}=k_{F}^{0}$,
and considering $|E_{F}-E_{k_{z}}^{1+}|\gg\hbar/\tau_{k_{x}k_{z}}^{1+}$,
\begin{eqnarray}
\sigma_{xx}^{0,1+} & \approx & \frac{e^{2}}{h}\left[\frac{|\hbar v_{0,1+}^{x}(k_{z})|^{2}}{|\hbar v_{F}^{0}|}\frac{1}{[(E_{F}-E_{k_{z}}^{1+})^{2}]}\right]_{k_{z}=k_{F}^{0}}\nonumber \\
 &  & \times\frac{1}{L_{y}}\int_{-L_{y}/2\ell_{B}^{2}}^{L_{y}/2\ell_{B}^{2}}\frac{dk_{x}}{2\pi}\left[\Lambda\frac{\hbar}{\tau_{k_{x}k_{z}}^{1+}}\right]_{k_{z}=k_{F}^{0}}.
\end{eqnarray}
This is evaluated in the numerical calculation. Because the $1+$
band does not cross the Fermi energy, $\tau_{k_{x}k_{z}}^{1+}$ is
mainly contributed by the virtual scattering processes with the $\nu=0$
band, i.e., $\tau_{k_{x}k_{z}}^{1+}\rightarrow\tau_{k_{x}k_{z}}^{1+\leftrightarrow0}$,
and
\begin{eqnarray}
\frac{\hbar}{\tau_{k_{x}k_{z}}^{1+\leftrightarrow0}} & = & 2\pi\sum_{k_{x}',k_{z}'}\langle|U_{k_{x}k_{z},k_{x}'k_{z}'}^{1+,0}|^{2}\rangle\Lambda\delta(E_{F}-E_{k_{z}'}^{0}).
\end{eqnarray}
Using Eq. (\ref{U-abcd-integral}) and $\sum_{i=1}^{2}1/\hbar|v_{0k_{z}'}^{z}|_{k_{z}'=k_{F}^{0,i}}=2/\hbar v_{F}^{0}$,
we then find Eqs. (\ref{tau-01}) and (\ref{sigma-xx-01+}). Similarly,
in the strong-field limit, one can obtain $\tau_{k_{z}k_{x}}^{1-\leftrightarrow0}$
and $\sigma_{xx}^{0,1-}$.


%

\end{document}